\RequirePackage{fix-cm}
\documentclass[smallextended]{svjour3}       % onecolumn (second format)
\smartqed  % flush right qed marks, e.g. at end of proof
\usepackage{graphicx}
\usepackage{enumerate}
\usepackage{subfigure}
\usepackage{amsmath}
\usepackage{amssymb}
\usepackage{array}
\usepackage{theorem}
\begin{document}

\title{Quantum Image Matching
\thanks{This work is supported by the National Natural Science Foundation of China under Grants No. 61502016, the Fundamental Research Funds for the Central Universities under Grants No. 2015JBM027, and the Graduate
Technology Fund of BJUT under Grants No. ykj-2015-11719.} }
%\subtitle{Do you have a subtitle?\\ If so, write it here}

%\titlerunning{Short form of title}        % if too long for running head

\author{Nan Jiang \and Yijie Dang \and Jian Wang}

%\authorrunning{Short form of author list} % if too long for running head

\institute{N. Jiang \at
              College of Computer Science, Beijing University of Technology, Beijing 100124, China \\
              College of Science, Purdue University, West Lafayette
              47905, USA \\
              Beijing Key Laboratory of Trusted Computing, Beijing 100124, China \\
              National Engineering Laboratory for Critical Technologies of
Information Security Classified Protection, Beijing 100124, China \\
           \and
Y.J. Dang \at College of Computer Science, Beijing University of
Technology, Beijing 100124, China\\
           \and
           J. Wang \at
              School of Computer and Information Technology, Beijing Jiaotong University, Beijing 100044, China \\
              College of Science, Purdue University, West Lafayette
              47905, USA \\
              \email{wangjian@bjtu.edu.cn}
             }

\date{Received: date / Accepted: date}
% The correct dates will be entered by the editor

\maketitle

\begin{abstract}
Quantum image processing (QIP) means the quantum based methods to
speed up image processing algorithms. Many quantum image processing
schemes claim that their efficiency are theoretically higher than
their corresponding classical schemes. However, most of them do not
consider the problem of measurement. As we all know, measurement
will lead to collapse. That is to say, executing the algorithm once,
users can only measure the final state one time. Therefore, if users
want to regain the results (the processed images), they must execute
the algorithms many times and then measure the final state many
times to get all the pixels' values. If the measurement process is
taken into account, whether or not the algorithms are really
efficient needs to be reconsidered. In this paper, we try to solve
the problem of measurement and give a quantum image matching
algorithm. Unlike most of the QIP algorithms, our scheme interests
only one pixel (the target pixel) instead of the whole image. It
modifies the probability of pixels based on Grover's algorithm to
make the target pixel to be measured with higher probability, and
the measurement step is executed only once. An example is given to
explain the algorithm more vividly. Complexity analysis indicates
that the quantum scheme's complexity is $O(2^{n})$ in
contradistinction to the classical scheme's complexity
$O(2^{2n+2m})$, where $m$ and $n$ are integers related to the size
of images.
% \PACS{PACS code1 \and PACS code2 \and more}
% \subclass{MSC code1 \and MSC code2 \and more}
\end{abstract}

\section{Introduction}

In 1982, Feynman proposed a novel computation model, named quantum
computers which can efficiently solve some problems that are
believed to be intractable on classical computers [1]. After that,
many researchers devote themselves into the research about quantum
computers. The latest research shows that quantum computers are 100
million times faster than classical computers [2].

The developing of quantum computer causes people's interest to study
quantum image processing (QIP) which refers to use quantum computers
to process images. Many researchers have proposed a number kinds of
QIP algorithms, such as geometric transformation [3-6], color
transformation [7-8], image scrambling [9-11], image segmentation
[12-14], feature extraction [15], quantum image watermark [16-23],
quantum image encryption [24-27], and quantum imaging [32]. Their
efficiency are theoretically higher than their corresponding
classical schemes.

However, most of them do not consider the problem of measurement.
Fig. 1(a) gives an example.
\begin{figure}[h]
  \centering
  \includegraphics[width=12cm,keepaspectratio]{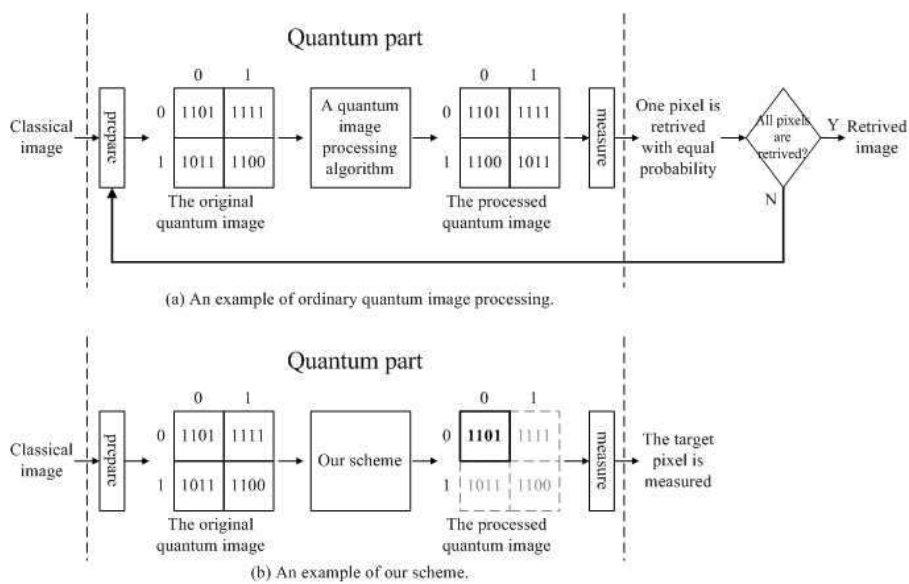}
  \caption{The difference between ordinary quantum image processing algorithms and our scheme.}
  \label{fig:1}
\end{figure}

No matter the image are the original one or the processed one, its
representation method is similar: a superposition state is used to
store the image and every pixel has the same probability to be
measured. Taking the processed image in Fig. 1(a) as an example, its
representation is
$$
|I\rangle=\frac{1}{2}|1101\rangle\otimes|00\rangle+\frac{1}{2}|1111\rangle\otimes|01\rangle+\frac{1}{2}|1100\rangle\otimes|10\rangle+\frac{1}{2}|1011\rangle\otimes|11\rangle.
$$

$|I\rangle$ is a superposition state. If we want to read the image
out, it must be measured. However, once the state is measured, it
will collapse to the non-superposition state that has been measured.
That is to say, once a pixel is measured, other pixels will
disappear. Like the image $|I\rangle$, it has 4 pixels and the
probabilities of a pixel being measured are all
$\left(\frac{1}{2}\right)^{2}=\frac{1}{4}$. If the pixel being
measured is $|1111\rangle\otimes|01\rangle$, $|I\rangle$ collapses
to $|1111\rangle\otimes|01\rangle$ and the other 3 pixels disappear.
How to measure the other pixels? The only method is performing the
quantum image processing again. Most fortunately, if a different
pixel is measured each time, by repeating the whole process 4 times,
the 4 pixels' values are regained. However, if we are not so
fortunate, it can not say the exact number of repetitions, maybe 10
times, maybe 100 times, and maybe never get all the 4 values.

Hence, in order to read out the result, users must execute the
quantum algorithm many times. Therefore, whether or not the
algorithms are really efficient needs to be reconsidered. In Ref.
[38-39], Mario Mastriani holds a similar view.

In this paper, we try to solve the problem of measurement and give a
quantum image matching algorithm (see Fig. 1(b)). This scheme
modifies the probability of pixels to make the target pixel to be
measured with higher probability and the whole scheme is executed
only once.

The rest of the paper is organized as follows. Sect. 2 gives related
works about quantum image matching. Sect. 3 introduces what image
matching is. Our scheme is discussed in Sect. 4 including the basic
ideas, the scheme steps, and the theoretical analysis. Sect. 5
analysis the network complexity. Sect. 6 gives the conclusion.

\section{Related works}

There have been several references that discuss the quantum image
matching.

Yang and \emph{etc} give a quantum gray-scale image matching scheme
in Ref. [34], in which the quantum template image is directly mapped
with quantum reference image, i.e., the quantum register
representing each corresponding pixel of the quantum template image
is subtracted from that of the quantum reference image by running a
quantum subtracter. According to the quantum measurement results,
the difference is saved and sum all the differences. Then compare
the sum with a Tolerance value. If the sum is smaller than the
Tolerance value, then quantum image matching succeeds. However, we
think that it has 3 shortcomings: (1) Although the title is
``quantum image matching'', in fact, the paper computes the
similarity between two quantum images instead of giving a quantum
image matching scheme. (2) What the quantum part of the scheme does
only is subtraction, which can be done more efficiently in classical
computers. (3) It needs to regain all the pixels in the different
image, i.e., it needs to be processed and measured many times.

A chaotic quantum-behaved particle swarm optimization based on
lateral inhibition (LI-CQPSO) is proposed in Ref. [35], which is
used to solve complicated image matching problems. In this work, the
proposed LI-CQPSO combines the techniques of chaos theory, quantum
and lateral inhibition. Chaos can guarantee the PSO escaping from
local best, quantum can make the traditional PSO with better
searching performance as well as having fewer parameters to control,
and lateral inhibition is applied to extract the edge of the images
by sharpening the spatial profile of excitation in response to a
localized stimulus. However, it is a classical algorithm and all
steps are done in classical computers. Quantum-behaved particle
swarm optimization (QPSO) is only motivated by concepts from quantum
mechanics, but is not a quantum algorithm.

Ref. [36] proposes to combine the shape context (SC) descriptor with
quantum genetic algorithms (QGA) to define a new shape matching and
retrieval method. The SC matching method is based on finding the
best correspondence between two point sets. However, it is suitable
for shape matching instead of natural images matching.

The algorithm proposed in Ref. [37] is similar to our work: change
the probability of the target pixel. However, as the authors say in
this paper, it is not a complete solution because that although the
probability of the target pixel is larger than the probability of
other pixels, it is still relatively small.

Unlike the above works, we will give a quantum image matching
solution which does not compute similarity, is not a classical
algorithm based on quantum concepts, and is not for shape matching.

\section{Image matching}

Image matching is the process of searching for a small image in a
big image. It is widely used in computer vision, military automatic
target recognition, face detection, manufacturing quality control,
visual positioning and tracking, and so on. Fig. 2 gives an example.

\begin{figure}[h]
  \centering
  \includegraphics[height=5cm,keepaspectratio]{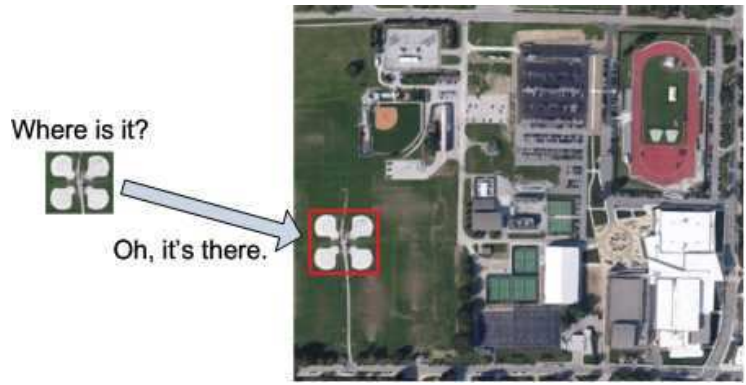}
  \caption{The application of image matching.}
  \label{fig:2}
\end{figure}

The simplest image matching method is the exhaustive search. Assume
that the size of the big image is $2^{n}\times2^{n}$ and the size of
the small image is $2^{m}\times2^{m}$, where $n>m$. For each
$2^{m}\times2^{m}$ image block in the big image, the method compares
all the pixels in the block with the corresponding ones in the small
image. If the block is the same as the small image, the method will
finish its work and output the block's location. Hence, the
complexity of the exhaustive search is $O(2^{2n+2m})$. It is
exponential and too high to bear in most application cases.

Therefore, researchers give other schemes to reduce the complexity.
Most of them take advantage of the features of images:
\begin{enumerate}[$\bullet$]

  \item Feature-based scheme compares the image features instead of
  the whole image pixels to match two images [28]. Color, edge, feature points and so on can all be used as features. Fig. 3 gives two examples. In general, no matter what kind
  of features, their data volume is obviously smaller than the original image (the
  pixels), which helps to reduce the complexity.

\begin{figure}[h]
  \centering
  \subfigure[Feature points (Only the red points are feature points.)]{
    \label{fig:subfig:a}
    \includegraphics[height=5cm,keepaspectratio]{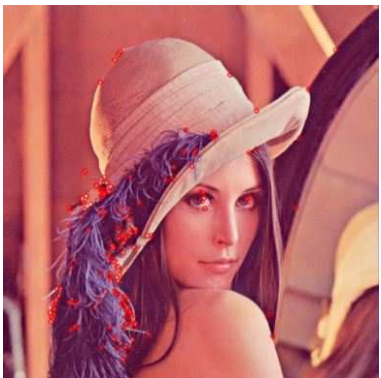}}
  \hspace{0.2in}
  \subfigure[Edge]{
    \label{fig:subfig:b}
    \includegraphics[height=5cm,keepaspectratio]{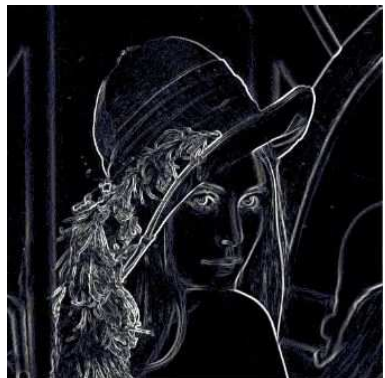}}
  \caption{Different kinds of features.}
  \label{fig:3}
\end{figure}

  \item Frequency-domain schemes find the transformation parameters for
  matching [29]. The commonly used transformations include the discrete cosine transform, the discrete Fourier
  transform, the discrete wavelet transform and so on. This method
  can not only reduce the complexity by cutting down the data volume
  to be compared, but also tolerate a certain degree of deformation
  caused by illumination, angle, distance and \emph{etc} (see Fig. 4).

\begin{figure}[h]
  \centering
  \subfigure[illumination]{
    \label{fig:subfig:a}
    \includegraphics[height=1.7cm,keepaspectratio]{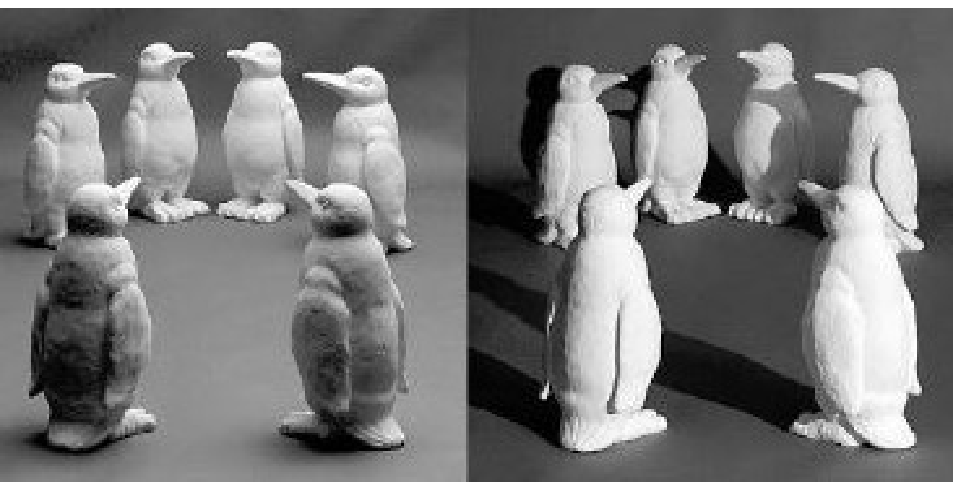}}
  \hspace{0.01in}
  \subfigure[angle]{
    \label{fig:subfig:b}
    \includegraphics[height=1.7cm,keepaspectratio]{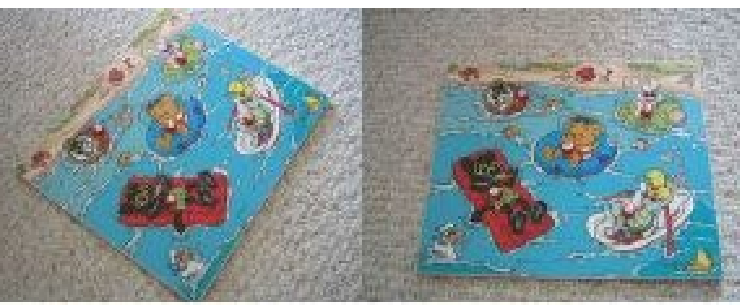}}
  \hspace{0.01in}
  \subfigure[distance]{
    \label{fig:subfig:c}
    \includegraphics[height=1.7cm,keepaspectratio]{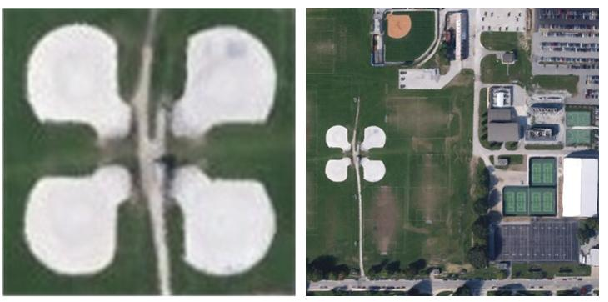}}
  \caption{Deformation caused by illumination, angle and distance, in which (a) and (b) are cited from [30].}
  \label{fig:4}
\end{figure}

  \item Interactive schemes provide tools to match the images manually which reduce user bias by performing certain key operations automatically while still relying on the user to guide the
  matching [29].

\end{enumerate}

Although these improved schemes enhance the performance of image
matching, it is still a difficult problem in image processing
because the efficiency is unsatisfactory yet.

In this paper, by taking advantage of the high computing ability of
quantum computers, we give a quantum image matching algorithm to
reduce the complexity.

\section{Quantum image matching algorithm}
\subsection{Basic ideas}
We use the generalized quantum image representation (GQIR) method
[6] to store quantum images in this paper. GQIR is developed from
the novel enhanced quantum representation (NEQR) [7] with the
difference that GQIR can represent arbitrary $H\times W$ images
instead of only $2^{n}\times2^{n}$ images. If a quantum image is
with size $2^{n}\times2^{n}$, its GQIR representation and NEQR
representation are the same. Readers please refer Ref. [6] about
GQIR's preparation and analysis. For the sake of simplicity, in this
paper, we only describe GQIR briefly.

An images can be seen as a function: $I(x,y)$. Hence, a GQIR image
is represented as an entangled state
\begin{equation}
|I(x,y)\rangle\otimes|xy\rangle
\end{equation}
Assume that the size of an image is $2^{n}\times2^{n}$ and
$I(x,y)\in\{0,1,\cdots,2^{q}-1\}$. Due to the equal role of every
pixels in an image, their probability are equivalent. Hence,
\begin{equation}
|xy\rangle=\frac{1}{\sqrt{2^{2n}}}\sum_{k=0}^{2^{2n}-1}|k\rangle=\frac{1}{2^{n}}\sum_{k=0}^{2^{2n}-1}|k\rangle
\end{equation}

In image matching, the inputs are two images: the big image $A$ with
size $2^{n}\times2^{n}$ and the small image $B$ with size
$2^{m}\times2^{m}$, where $n>m$. The matching algorithm's output is
the location $(x,y)$ of the upper left pixel of $B$ in $A$. Fig. 5
gives a sketch map.
\begin{figure}[h]
  \centering
  \includegraphics[height=5cm,keepaspectratio]{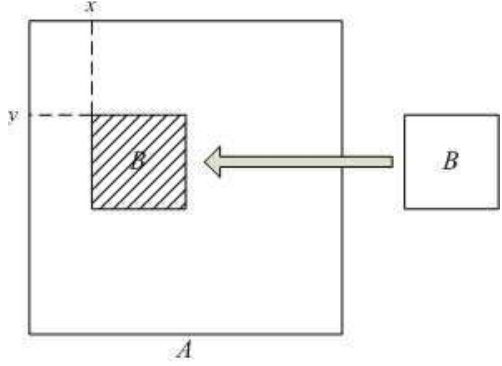}
  \caption{The output of image matching. If the algorithm finds out $B$ in the shaded area in $A$, it will output the upper left location $(x,y)$.}
  \label{fig:5}
\end{figure}

In order to get $(x,y)$ in measurement, the basic idea of our
solution is to increase the probability of $(x,y)$ and reduce the
probability of other pixels. As early as 1996, Grover had provided
the idea [31]. In this paper, we use it into quantum image matching.

\subsection{Algorithm}

According to the previous presentation, the two input images $A$ and
$B$ are
\begin{equation}
|A\rangle=\frac{1}{2^{n}}\sum_{k_{A}=0}^{2^{2n}-1}|I_{A}(k_{A})\rangle\otimes|k_{A}\rangle
\end{equation}
\begin{equation}
|B\rangle=\frac{1}{2^{m}}\sum_{k_{B}=0}^{2^{2m}-1}|I_{B}(k_{B})\rangle\otimes|k_{B}\rangle
\end{equation}
Additionally, two auxiliary qubit $|f\rangle=|0\rangle$ and
$|g\rangle=\frac{1}{\sqrt{2}}(|0\rangle-|1\rangle)$ are needed.
Hence, the initial state of the algorithm is
\begin{equation}
\begin{split}
\Psi_{0}&=|g\rangle\otimes|f\rangle\otimes|A\rangle\otimes|B\rangle\\
&=|g\rangle\otimes\left(\frac{1}{2^{n+m}}\sum_{k_{A}=0}^{2^{2n}-1}\sum_{k_{B}=0}^{2^{2m}-1}|0\rangle\otimes|I_{A}(k_{A})\rangle\otimes|k_{A}\rangle\otimes|I_{B}(k_{B})\rangle\otimes|k_{B}\rangle\right)
\end{split}
\end{equation}

\begin{enumerate}[Step 1]
  \item Find out the matched area, i.e., the pixels that $I_{A}=I_{B}$.
\end{enumerate}

Due to $I_{A},I_{B}\in\{0,1,\cdots,2^{q}-1\}$, $q$ qubits are used
to store $|I_{A}\rangle$ and $|I_{B}\rangle$.
\begin{equation}
|I_{T}(k_{T})\rangle=|I^{q-1}_{T}(k_{T})I^{q-2}_{T}(k_{T})\cdots
I^{0}_{T}(k_{T})\rangle
\end{equation}
where, $T\in\{A,B\}$ and $I^{i}_{T}(k_{T})\in\{0,1\}$,
$i=q-1,q-2,\cdots,0$.

Define
\begin{equation}
U_{1}^{i}:|I^{i}_{A},I^{i}_{B}\rangle\rightarrow|I^{i}_{A}\oplus
I^{i}_{B},I^{i}_{B}\rangle
\end{equation}
That is to say, $U_{1}^{i}$ is a CNOT gate: if
$I^{i}_{A}=I^{i}_{B}$, $I^{i}_{A}$ is changed to 0; otherwise,
$I^{i}_{A}$ is changed to 1. Hence, by acting
\begin{equation}
U_{1}=\bigotimes_{i=q-1}^{0}U_{1}^{i}
\end{equation}
to state $\Psi_{0}$, we can get
\begin{equation}
\begin{split}
\Psi_{1}&=U_{1}(\Psi_{0})\\
&=|g\rangle\otimes\left(\frac{1}{2^{n+m}}\sum_{k_{A}}\sum_{k_{B}}|0\rangle\otimes[\otimes_{i=q-1}^{0}|I^{i}_{A}(k_{A})\oplus
I^{i}_{B}(k_{B})\rangle]\otimes|k_{A}\rangle\otimes|I_{B}(k_{B})\rangle\otimes|k_{B}\rangle\right)\\
&=|g\rangle\otimes\left(\frac{1}{2^{n+m}}\sum_{k_{A}}\sum_{k_{B}}|0\rangle\otimes|I_{A}(k_{A})\oplus
I_{B}(k_{B})\rangle\otimes|k_{A}\rangle\otimes|I_{B}(k_{B})\rangle\otimes|k_{B}\rangle\right)
\end{split}
\end{equation}
For the sake of simplicity, when no ambiguity is possible, we still
use $I_{A}(k_{A})$ to substitute $I_{A}(k_{A})\oplus I_{B}(k_{B})$.
Therefore
\begin{equation}
\begin{split}
\Psi_{1}=&|g\rangle\otimes\left(\frac{1}{2^{n+m}}\sum_{k_{A}}\sum_{k_{B}}|0\rangle\otimes|I_{A}(k_{A})\rangle\otimes|k_{A}\rangle\otimes|I_{B}(k_{B})\rangle\otimes|k_{B}\rangle\right)\\
=&|g\rangle\otimes\left(\frac{1}{2^{n+m}}\left[\sum_{k_{A},k_{B},\text{ and }I_{A}=I_{B}}|0\rangle\otimes|0\rangle^{\otimes q}\otimes|k_{A}\rangle\otimes|I_{B}(k_{B})\rangle\otimes|k_{B}\rangle\right.\right.\\
&\left.\left.+\sum_{k_{A},k_{B},\text{ and }I_{A}\neq
I_{B}}|0\rangle\otimes|I_{A}(k_{A})\rangle\otimes|k_{A}\rangle\otimes|I_{B}(k_{B})\rangle\otimes|k_{B}\rangle\right]\right)
\end{split}
\end{equation}

Therefore, in image $A$, all the pixels that $I_{A}=I_{B}$, i.e., in
state $\Psi_{1}$, all the pixels that
$|I_{A}\rangle=|0\rangle^{\otimes q}$, consist the matched area.

\begin{enumerate}[Step 2]
  \item Find out the upper left corner of the matched area.
\end{enumerate}

Although the matched area is found out, the upper left corner of it
is still unknown. However, in image $B$, the upper left corner is
fixed: it is pixel $(0,0)$. Hence, in
$\Psi_{1}=|g\rangle\otimes\left(\frac{1}{2^{n+m}}\sum_{k_{A}}\sum_{k_{B}}|0\rangle\otimes|I_{A}(k_{A})\rangle\otimes|k_{A}\rangle\otimes|I_{B}(k_{B})\rangle\otimes|k_{B}\rangle\right)$,
when $|I_{A}(k_{A})\rangle=|0\rangle^{\otimes q}$ and
$|k_{B}\rangle=|0\rangle^{\otimes m}$, their corresponding
$|k_{A}\rangle$ is the quantum image matching algorithm's output. We
call it as $|k_{A_{0}}\rangle$. In Step 2, a transform $U_{2}$ is
used to change the state of the auxiliary qubit $|f\rangle$.
\begin{equation}
U_{2}:|f=0\rangle\rightarrow|f=1\rangle\text{, if
}|I_{A}(k_{A})\rangle=|0\rangle^{\otimes q}\text{ and
}|k_{B}\rangle=|0\rangle^{\otimes m}
\end{equation}
where, $U_{2}$ is a $(q+m)$-CNOT gate (a CNOT gate with $(q+m)$
control qubits).

Act $U_{2}$ to $\Psi_{1}$
\begin{equation}
\begin{split}
\Psi_{2}=&U_{2}(\Psi_{1})\\
=&|g\rangle\otimes\left(\frac{1}{2^{n+m}}\left[|1\rangle\otimes|0\rangle^{\otimes q}\otimes|k_{A_{0}}\rangle\otimes|I_{B}(k_{B})\rangle\otimes|0\rangle^{\otimes 2m}\right.\right.\\
&\left.\left.+\sum_{I_{A}\neq|0\rangle^{\otimes q}\text{ or
}|k_{B}\rangle\neq|0\rangle^{\otimes
m}}|0\rangle\otimes|I_{A}(k_{A})\rangle\otimes|k_{A}\rangle\otimes|I_{B}(k_{B})\rangle\otimes|k_{B}\rangle\right]\right)
\end{split}
\end{equation}

That is to say, in the superposition state $\Psi_{2}$, the basis
that $|f\rangle=|1\rangle$ corresponds to the upper left corner of
the matched area.

\begin{enumerate}[Step 3]
  \item Change the probability of subspace $|k_{A}\rangle$.
\end{enumerate}

In the first two steps, the upper left corner of the matched area
has been found out according to the color information
$|I_{A}\rangle$ and $|I_{B}\rangle$ and location information
$|k_{B}\rangle$. $|f\rangle$ is the deliverable of the first two
steps. Hence, The three information ($|I_{A}\rangle$,
$|I_{B}\rangle$ and $|k_{B}\rangle$) has no use in the following and
$|f\rangle$ substitutes them.

Therefore, in this step, we only work on the subspace
$|k_{A}\rangle$ and increase the probability of $|k_{A_{0}}\rangle$
based on Grover's algorithm. Hence, the initial state of Step 3 is
\begin{equation}
\begin{split}
\Psi_{30}=&|k_{A}\rangle\otimes|g\rangle\otimes|f\rangle\\
=&\frac{1}{2^{n}\sqrt{2}}\left[|k_{A_{0}}\rangle\otimes(|0\rangle-|1\rangle)\otimes|1\rangle+\sum_{f=0}|k_{A}\rangle\otimes(|0\rangle-|1\rangle)\otimes|f\rangle\right]
\end{split}
\end{equation}

\begin{enumerate}[Step 3.1]
  \item Rotate the phase of $|k_{A_{0}}\rangle$ by $\pi$ radians with the help of the auxiliary qubit $|g\rangle$ [33].

Define
\begin{equation}
U_{31}:|k_{A},g,f\rangle\rightarrow|k_{A},g\oplus f,f\rangle
\end{equation}
Obviously, $U_{31}$ is a CNOT gate. Then
\begin{equation}
\begin{split}
\Psi_{31}=&U_{31}(\Psi_{30})\\
=&\frac{1}{2^{n}\sqrt{2}}\left[|k_{A_{0}}\rangle\otimes(|1\rangle-|0\rangle)\otimes|1\rangle+\sum_{f=0}|k_{A}\rangle\otimes(|0\rangle-|1\rangle)\otimes|f\rangle\right]\\
=&-\frac{1}{2^{n}}|k_{A_{0}}\rangle\otimes|g\rangle\otimes|f\rangle+\frac{1}{2^{n}}\sum_{f=0}|k_{A}\rangle\otimes|g\rangle\otimes|f\rangle
\end{split}
\end{equation}
Hence, in the subspace $|k_{A}\rangle$,
\begin{equation}
\Psi_{31}=\left(\frac{1}{2^{n}},\cdots,\frac{1}{2^{n}},-\frac{1}{2^{n}},\frac{1}{2^{n}},\cdots,\frac{1}{2^{n}}\right)^{\text{T}}
\end{equation}
That is to say, the function of Step 3.1 is to change the
coefficient of $k_{A_{0}}$ to a negative.

  \item Increase the probability of $|k_{A_{0}}\rangle$ and decrease the
  probability of other $|k_{A}\rangle$.

Apply the diffusion transform $D$ on $\Psi_{31}$ which is defined by
the matrix $D$ as follows [31]:
\begin{equation}
D_{ij}=\frac{2}{2^{2n}}\text{ if }i\neq j,\text{ and
}D_{ii}=\frac{2}{2^{2n}}-1
\end{equation}
This diffusion transform $D$ can be implemented as $D=WRW$, where
$R$ is the rotation matrix and $W$ is the Walsh-Hadamard Transform
Matrix are defined as follows:
\begin{equation}
R_{ij}=0\text{ if }i\neq j,\text{ and }R_{00}=1,\text{ and
}R_{ii}=-1\text{ if }i\neq0
\end{equation}
\begin{equation}
W_{ij}=\frac{1}{2^{n}}(-1)^{\overline{i}\cdot\overline{j}}
\end{equation}
where $\overline{i}$ is the binary representation of $i$, and
$\overline{i}\cdot\overline{j}$ denotes the bitwise dot product of
the two strings $\overline{i}$ and $\overline{j}$.

After Step 3.2, the probability of $|k_{A_{0}}\rangle$ is increased
and the probability of other pixels is decreased.

\item Repeat the unitary operations Step 3.1 and 3.2
\begin{equation}
\hat{i}\approx0.7962a-0.6057
\end{equation}
times, where $a=2^{n}$ (Eq. (20) will be proved in Section 4.3). In
the subspace $|k_{A}\rangle$, the final state of this step is
\begin{equation}
\Psi_{33}=\left(t_{\hat{i}},\cdots,t_{\hat{i}},t_{\hat{i}0},t_{\hat{i}},\cdots,t_{\hat{i}}\right)^{\text{T}}
\end{equation}
where, $t_{\hat{i}0}$ is the corresponding element of
$|k_{A_{0}}\rangle$ and $t_{\hat{i}}$ is the corresponding element
of other pixels.
\end{enumerate}

\begin{enumerate}[Step 4]
  \item Measurement.
\end{enumerate}

Use projective measurements to measure state $\Psi_{33}$ and define
measurement operators as
\begin{equation}
P_{j}=|j\rangle\langle j|,\ j=0,1,\cdots,2^{2n-1}
\end{equation}
Hence, the probability that the basis $|j\rangle$ being measured is
\begin{equation}
\langle\Psi_{33}|P_{j}|\Psi_{33}\rangle=(t_{j})^{2}
\end{equation}
where, $t_{j}$ is the $j$th element in $\Psi_{33}$,
$j=0,1,\cdots,2^{2n-1}$. Hence, the basis $|k_{A_{0}}\rangle$ (i.e.,
the upper left corner of the matched area) is measured with
probability $t_{\hat{i}0}^{2}$ and
\begin{equation}
t_{\hat{i}0}^{2}\geq(0.9194+0.0567a^{-1}+0.2302a^{-2}-0.0336a^{-3})^{2}.
\end{equation}
(Eq. (24) will be proved in Section 4.3)

Fig. 6 gives the circuit of quantum image matching.
\begin{figure}[h]
  \centering
  \includegraphics[width=11cm,keepaspectratio]{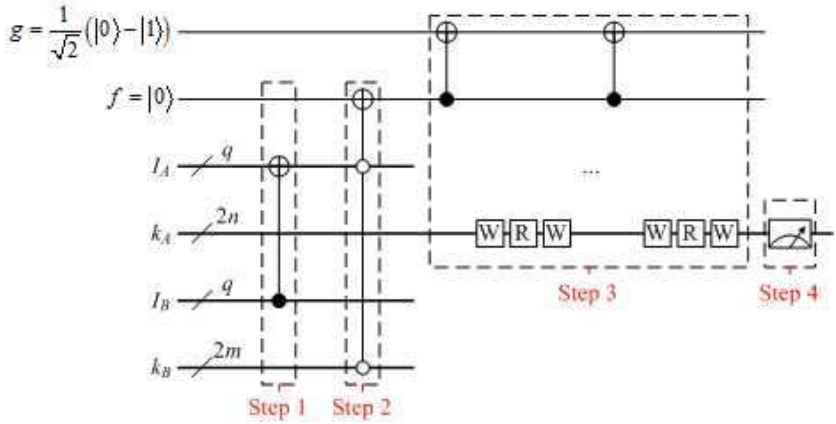}
  \caption{The circuit.}
  \label{fig:6}
\end{figure}

\subsection{Lemmas and theorems}

In Section 4.2, there are two key variables: $\hat{i}$ and
$t_{\hat{i}0}$. In this section, we will prove their values, i.e.,
Eq. (20) and (24).

\begin{lemma}%Lemma 1
The output of Step 3.2 after the $i$th iteration is in the form of
\begin{equation}
T_{i}=\left(t_{i},\cdots,t_{i},t_{i0},t_{i},\cdots,t_{i}\right)^{\text{T}}
\end{equation}
and
\begin{equation}
\begin{split}
t_{(i+1)0}=&-\frac{2t_{i0}}{a^{2}}-\frac{2t_{i}}{a^{2}}+2t_{i}+t_{i0}\\
t_{i+1}=&-\frac{2t_{i0}}{a^{2}}-\frac{2t_{i}}{a^{2}}+t_{i}
\end{split}
\end{equation}
where $a=2^{n}$. That is to say, after the $i$th iteration, the
output $T_{i}$ only has two values: $t_{i}$ and $t_{i0}$ and
$t_{i0}^{2}$ is the probability of $|k_{A_{0}}\rangle$ being
measured and $t_{i}^{2}$ is the probability of other $|k_{A}\rangle$
being measured.
\end{lemma}
\begin{proof}
According to Eq. (17),
\begin{equation}
D\equiv-I+2P
\end{equation}
where $I$ is the identity matrix and $P$ is a projection matrix with
$P_{ij}=\frac{1}{2^{2n}}$ for all $i,j$. Hence, for an arbitrary
$2^{2n}\times1$ matrix $S$
\begin{equation}
\begin{split}
&DS=D\left(
\begin{array}{c}
s_{0}\\
s_{1}\\
\cdots\\
s_{2^{2n}-1}\\
\end{array}
\right) =(2P-I)\left(
\begin{array}{c}
s_{0}\\
s_{1}\\
\cdots\\
s_{2^{2n}-1}\\
\end{array}
\right)\\
=&\left(
\begin{array}{c}
2\cdot\frac{s_{0}+s_{1}+\cdots+s_{2^{2n}-1}}{2^{2n}}-s_{0}\\
2\cdot\frac{s_{0}+s_{1}+\cdots+s_{2^{2n}-1}}{2^{2n}}-s_{1}\\
\cdots\\
2\cdot\frac{s_{0}+s_{1}+\cdots+s_{2^{2n}-1}}{2^{2n}}-s_{2^{2n}-1}\\
\end{array}
\right)=\left(
\begin{array}{c}
2\overline{s}-s_{0}\\
2\overline{s}-s_{1}\\
\cdots\\
2\overline{s}-s_{2^{2n}-1}\\
\end{array}
\right)
\end{split}
\end{equation}
where $\overline{s}=\frac{s_{0}+s_{1}+\cdots+s_{2^{2n}-1}}{2^{2n}}$
is the average value of $s_{0},s_{1},\cdots,s_{2^{2n}-1}$. From Eq.
(28), it is obviously that if $s_{i}=s_{j}$, then
$2\overline{s}-s_{i}=2\overline{s}-s_{j}$. Due to the input of Step
3.2 in the first iteration is
$\Psi_{31}=\left(\frac{1}{2^{n}},\cdots,\frac{1}{2^{n}},-\frac{1}{2^{n}},\frac{1}{2^{n}},\cdots,\frac{1}{2^{n}}\right)^{\text{T}}$
which only has two values: $\frac{1}{2^{n}}$ and $-\frac{1}{2^{n}}$,
i.e., $t_{00}=t_{0}=\frac{1}{2^{n}}=\frac{1}{a}$, hence
$$
T_{1}=\left(t_{1},\cdots,t_{1},t_{10},t_{1},\cdots,t_{1}\right)^{\text{T}}
$$
and if we use $\overline{t}_{i}$ to respect the average value in the
$i$th iteration,
$$
\overline{t}_{1}=\left. \left((a^{2}-1)\cdot t_{0}-t_{00}\right)
\middle / a^{2} \right.=\left.
\left((a^{2}-1)\cdot\frac{1}{a}-\frac{1}{a}\right) \middle / a^{2}
\right.=\frac{1}{a}-\frac{2}{a^{3}}
$$
then
$$
t_{10}=2\overline{t}_{1}+t_{00}=\frac{3}{a}-\frac{4}{a^{3}}=-\frac{2t_{00}}{a^{2}}-\frac{2t_{0}}{a^{2}}+2t_{0}+t_{00}
$$
$$
t_{1}=2\overline{t}_{1}-t_{0}=\frac{1}{a}-\frac{4}{a^{3}}=-\frac{2t_{00}}{a^{2}}-\frac{2t_{0}}{a^{2}}+t_{0}
$$

By assuming that after the $l$th iteration,
$$
T_{l}=\left(t_{l},\cdots,t_{l},t_{l0},t_{l},\cdots,t_{l}\right)^{\text{T}},
$$
and
\begin{equation*}
\begin{split}
t_{l0}=&-\frac{2t_{(l-1)0}}{a^{2}}-\frac{2t_{(l-1)}}{a^{2}}+2t_{(l-1)}+t_{(l-1)0}\\
t_{l}=&-\frac{2t_{(l-1)0}}{a^{2}}-\frac{2t_{(l-1)}}{a^{2}}+t_{(l-1)}
\end{split}
\end{equation*}
according to the function of Step 3.1, the input of Step 3.2 in the
$(l+1)$th iteration will be
$$
\left(t_{l},\cdots,t_{l},-t_{l0},t_{l},\cdots,t_{l}\right)^{\text{T}}.
$$
Hence, after the $(l+1)$th iteration,
$$
T_{l+1}=\left(t_{l+1},\cdots,t_{l+1},t_{(l+1)0},t_{l+1},\cdots,t_{l+1}\right)^{\text{T}}.
$$
and
\begin{equation*}
\begin{split}
\overline{t}_{l+1}=&\left. \left((a^{2}-1)\cdot t_{l}-t_{l0}\right)
\middle / a^{2} \right.=-\frac{t_{l0}}{a^{2}}-\frac{t_{l}}{a^{2}}+t_{l}\\
t_{(l+1)0}=&2\overline{t}_{l+1}+t_{l0}=-\frac{2t_{l0}}{a^{2}}-\frac{2t_{l}}{a^{2}}+2t_{l}+t_{l0}\\
t_{l+1}=&2\overline{t}_{l+1}-t_{l}=-\frac{2t_{l0}}{a^{2}}-\frac{2t_{l}}{a^{2}}+t_{l}
\end{split}
\end{equation*}

According to mathematical induction, the lemma is proved.
\end{proof}

\begin{lemma}%Lemma 2
\begin{equation}
t_{i0}=\frac{2i+1}{a}+\sum_{j=2,j=j+1}^{i}(-1)^{j-1}\frac{f(i^{2j-1})}{a^{2j-1}}+(-1)^{i}\frac{2^{2i}}{a^{2i+1}}
\end{equation}
\begin{equation}
t_{i}=\frac{1}{a}+\sum_{j=2,j=j+1}^{i}(-1)^{j-1}\frac{f(i^{2j-2})}{a^{2j-1}}+(-1)^{i}\frac{2^{2i}}{a^{2i+1}}
\end{equation}
where, $f(i^{d})$ represents a $d$-order polynomial about $i$, and
in Eq. (29), if $j=2$,
$$
(-1)^{j-1}\frac{f(i^{2j-1})}{a^{2j-1}}=-\frac{\frac{2}{3}i(i+1)(2i+1)}{a^{3}},
$$
in Eq. (30), if $j=2$,
$$
(-1)^{j-1}\frac{f(i^{2j-2})}{a^{2j-1}}=-\frac{2i(i+1)}{a^{3}},
$$
and in Eq. (30), if $j=3$,
$$
(-1)^{j-1}\frac{f(i^{2j-2})}{a^{2j-1}}=\frac{\frac{2}{3}i(i+2)(i^{2}-1)}{a^{5}}.
$$
\end{lemma}
\begin{proof}
When $i=1$,
\begin{equation*}
\begin{split}
t_{10}=&-\frac{2t_{00}}{a^{2}}-\frac{2t_{0}}{a^{2}}+2t_{0}+t_{00}=\frac{3}{a}-\frac{4}{a^{3}}=\frac{2\times1+1}{a}+(-1)^{1}\frac{2^{2\times1}}{a^{2\times1+1}}\\
t_{1}=&-\frac{2t_{00}}{a^{2}}-\frac{2t_{0}}{a^{2}}+t_{0}=\frac{1}{a}-\frac{4}{a^{3}}=\frac{1}{a}+(-1)^{1}\frac{2^{2\times1}}{a^{2\times1+1}}
\end{split}
\end{equation*}

When $i=2$,
\begin{equation*}
\begin{split}
t_{20}=&-\frac{2t_{10}}{a^{2}}-\frac{2t_{1}}{a^{2}}+2t_{1}+t_{10}=\frac{5}{a}-\frac{20}{a^{3}}+\frac{16}{a^{5}}\\
=&\frac{2\times2+1}{a}-\frac{\frac{2}{3}\times2(2+1)(2\times2+1)}{a^{3}}+(-1)^{2}\frac{2^{2\times2}}{a^{2\times2+1}}\\
t_{2}=&-\frac{2t_{10}}{a^{2}}-\frac{2t_{1}}{a^{2}}+t_{1}=\frac{1}{a}-\frac{12}{a^{3}}+\frac{16}{a^{5}}\\
=&\frac{1}{a}-\frac{2\times2(2+1)}{a^{3}}+(-1)^{2}\frac{2^{2\times2}}{a^{2\times2+1}}
\end{split}
\end{equation*}

When $i=3$,
\begin{equation*}
\begin{split}
t_{30}=&-\frac{2t_{20}}{a^{2}}-\frac{2t_{2}}{a^{2}}+2t_{2}+t_{20}=\frac{7}{a}-\frac{56}{a^{3}}+\frac{112}{a^{5}}-\frac{64}{a^{7}}\\
=&\frac{2\times3+1}{a}-\frac{\frac{2}{3}\times3(3+1)(2\times3+1)}{a^{3}}+\frac{112}{a^{5}}+(-1)^{3}\frac{2^{2\times3}}{a^{2\times3+1}}\\
t_{3}=&-\frac{2t_{20}}{a^{2}}-\frac{2t_{2}}{a^{2}}+t_{2}=\frac{1}{a}-\frac{24}{a^{3}}+\frac{80}{a^{5}}-\frac{64}{a^{7}}\\
=&\frac{1}{a}-\frac{2\times3(3+1)}{a^{3}}+\frac{\frac{2}{3}\times3(3+2)(3^{2}-1)}{a^{5}}+(-1)^{3}\frac{2^{2\times3}}{a^{2\times3+1}}
\end{split}
\end{equation*}

By assuming that after the $l$th iteration,
\begin{equation*}
t_{l0}=\frac{2l+1}{a}-\frac{\frac{2}{3}l(l+1)(2l+1)}{a^{3}}+\frac{f(l^{5})}{a^{5}}+\cdots+(-1)^{l-1}\frac{f(l^{2l-1})}{a^{2l-1}}+(-1)^{l}\frac{2^{2l}}{a^{2l+1}}
\end{equation*}
\begin{equation*}
t_{l}=\frac{1}{a}-\frac{2l(l+1)}{a^{3}}+\frac{\frac{2}{3}l(l+2)(l^{2}-1)}{a^{5}}+\cdots+(-1)^{l-1}\frac{f(l^{2l-2})}{a^{2l-1}}+(-1)^{l}\frac{2^{2l}}{a^{2l+1}}
\end{equation*}
then after the $(l+1)$th iteration,
\begin{equation*}
\begin{split}
t_{(l+1)0}=&-\frac{2t_{l0}}{a^{2}}-\frac{2t_{l}}{a^{2}}+2t_{l}+t_{l0}\\
=&\frac{2(l+1)+1}{a}-\frac{\frac{2}{3}(l+1)((l+1)+1)(2(l+1)+1)}{a^{3}}+\frac{f((l+1)^{5})}{a^{5}}+\cdots\\
&+(-1)^{(l+1)-1}\frac{f((l+1)^{2(l+1)-1})}{a^{2(l+1)-1}}+(-1)^{(l+1)}\frac{2^{2(l+1)}}{a^{2(l+1)+1}}\\
t_{l+1}=&-\frac{2t_{l0}}{a^{2}}-\frac{2t_{l}}{a^{2}}+t_{l}\\
=&\frac{1}{a}-\frac{2(l+1)((l+1)+1)}{a^{3}}+\frac{\frac{2}{3}(l+1)((l+1)+2)((l+1)^{2}-1)}{a^{5}}+\cdots\\
&+(-1)^{(l+1)-1}\frac{f((l+1)^{2(l+1)-2})}{a^{2(l+1)-1}}+(-1)^{(l+1)}\frac{2^{2(l+1)}}{a^{2(l+1)+1}}
\end{split}
\end{equation*}

According to mathematical induction, the lemma is proved.
\end{proof}

In fact, the first item in $t_{i0}$ (Eq. (29)) also can be denoted
as $\frac{f(i^{1})}{a^{1}}$ and the first item in $t_{i}$ (Eq. (30))
also can be denoted as $\frac{f(i^{0})}{a^{1}}$.

\begin{lemma}%Lemma 3
In $t_{i0}$ (i.e., Eq. (29)),
\begin{equation}
2^{2i}<a^{2}f(i^{2i-1})
\end{equation}
and in $t_{i}$ (i.e., Eq. (30)),
\begin{equation}
2^{2i}<a^{2}f(i^{2i-2})
\end{equation}
\end{lemma}
\begin{proof}
In Eq. (29) and (30), the smallest $i$ that make $f(i^{2j-1})$ and
$f(i^{2j-2})$ exist is 2. According to Lemma 1 and because
$t_{00}=t_{0}=\frac{1}{a}$
\begin{equation*}
\begin{split}
t_{20}&=\frac{5}{a}-\frac{20}{a^{3}}+\frac{16}{a^{5}}\\
t_{2}&=\frac{1}{a}-\frac{12}{a^{3}}+\frac{16}{a^{5}}
\end{split}
\end{equation*}
According to the property of Grover's algorithm, the number of
iterations $i$ is no more than $a$, i.e., $i\leq a$. Hence, when
$i=2$, $a\geq2$. As a consequence,
$$
16<20\times4\leq20a^{2}\ \text{ and }\ 16<12\times4\leq12a^{2}
$$

When $i=3$, $a\geq3$ and
\begin{equation*}
\begin{split}
t_{30}&=\frac{7}{a}-\frac{56}{a^{3}}+\frac{112}{a^{5}}-\frac{64}{a^{7}}\\
t_{3}&=\frac{1}{a}-\frac{24}{a^{3}}+\frac{80}{a^{5}}-\frac{64}{a^{7}}
\end{split}
\end{equation*}
As a consequence,
$$
64<112\times9\leq112a^{2}\ \text{ and }\ 64<80\times9\leq80a^{2}
$$

By assuming that after the $l$th ($l\geq3$) iteration, In $t_{l0}$,
$$
2^{2l}<a^{2}f(l^{2l-1})
$$
$$
2^{2l}<a^{2}f(l^{2l-2})
$$
then after the ($l+1$)th iteration, according to Lemma 1, in
$t_{l0}$,
\begin{equation*}
\begin{split}
&2^{2(l+1)}-a^{2}f((l+1)^{2(l+1)-1})\\
=&4\times2^{2l}-a^{2}\left(\frac{2}{a^{2}}f(l^{2l-1})+\frac{2}{a^{2}}f(l^{2l-2})+2\times2^{2l}+2^{2l}\right)\\
=&4\times2^{2l}-2f(l^{2l-1})-2f(l^{2l-2})-3a^{2}2^{2l}\\
<&4\times2^{2l}-2\frac{2^{2l}}{a^{2}}-2\frac{2^{2l}}{a^{2}}-3a^{2}2^{2l}\\
=&-\frac{2^{2l}}{a^{2}}\left(2a^{4}+(a^{2}-2)^{2}\right)<0
\end{split}
\end{equation*}
and in $t_{l}$,
\begin{equation*}
\begin{split}
&2^{2(l+1)}-a^{2}f((l+1)^{2(l+1)-2})\\
=&4\times2^{2l}-a^{2}\left(\frac{2}{a^{2}}f(l^{2l-1})+\frac{2}{a^{2}}f(l^{2l-2})+2^{2l}\right)\\
=&4\times2^{2l}-2f(l^{2l-1})-2f(l^{2l-2})-a^{2}2^{2l}\\
<&4\times2^{2l}-2\frac{2^{2l}}{a^{2}}-2\frac{2^{2l}}{a^{2}}-a^{2}2^{2l}\\
=&-\frac{2^{2l}}{a^{2}}(a^{2}-2)^{2}<0
\end{split}
\end{equation*}

According to mathematical induction, the lemma is proved.
\end{proof}

\begin{lemma}%Lemma 4
In $t_{i0}$ (i.e., Eq. (29)),
\begin{equation}
\frac{f(i^{2j+1})}{f(i^{2j-1})}<a^{2},\ j=2,3,\cdots,i-1,
\end{equation}
and in $t_{i}$ (i.e., Eq. (30)),
\begin{equation}
\frac{f(i^{2j})}{f(i^{2j-2})}<a^{2},\ j=2,3,\cdots,i-1,
\end{equation}
\end{lemma}
\begin{proof}
The smallest $i$ that make $f(i^{2j+1})$, $f(i^{2j-1})$, $f(i^{2j})$
and $f(i^{2j-2})$ exist is 3. According to Lemma 1 and because
$t_{00}=t_{0}=\frac{1}{a}$
\begin{equation*}
\begin{split}
t_{30}&=\frac{7}{a}-\frac{56}{a^{3}}+\frac{112}{a^{5}}-\frac{64}{a^{7}}\\
t_{3}&=\frac{1}{a}-\frac{24}{a^{3}}+\frac{80}{a^{5}}-\frac{64}{a^{7}}
\end{split}
\end{equation*}
According to the property of Grover's algorithm, the number of
iterations $i$ is no more than $a$, i.e., $i\leq a$. Hence, when
$i=3$, $a\geq3$. As a consequence,
$$
\frac{112}{56}=2<9\leq a^{2}\ \text{ and }\ \frac{80}{24}=3.33<9\leq
a^{2}
$$

When $i=4$, $a\geq4$ and
\begin{equation*}
\begin{split}
t_{40}&=\frac{9}{a}-\frac{120}{a^{3}}+\frac{432}{a^{5}}-\frac{576}{a^{7}}+\frac{256}{a^{9}}\\
t_{4}&=\frac{1}{a}-\frac{40}{a^{3}}+\frac{240}{a^{5}}-\frac{448}{a^{7}}+\frac{256}{a^{9}}
\end{split}
\end{equation*}
As a consequence,
$$
\frac{432}{120}=3.6<16\leq a^{2},\ \frac{576}{432}=1.33<16\leq a^{2}
$$
and
$$
\frac{240}{40}=6<16\leq a^{2},\ \frac{448}{240}=1.87<16\leq a^{2}
$$

By assuming that after the $l$th ($l\geq4$) iteration, In $t_{l0}$,
$$
\frac{f(l^{2j+1})}{f(l^{2j-1})}<a^{2},\ j=2,3,\cdots,l-1,
$$
and in $t_{l}$,
$$
\frac{f(l^{2j})}{f(l^{2j-2})}<a^{2},\ j=2,3,\cdots,l-1,
$$
then after the ($l+1$)th iteration, according to Lemma 1, in
$t_{l0}$,
\begin{equation*}
f((l+1)^{2j-1})=\left\{
\begin{array}{ll}
\frac{2}{a^{2}}(2l+1)+\frac{2}{a^{2}}+2\cdot2l(l+1)+\frac{2}{3}l(l+1)(2l+1)&j=2\\
\frac{2}{a^{2}}f(l^{2j-3})+\frac{2}{a^{2}}f(l^{2j-4})+2f(l^{2j-2})+f(l^{2j-1})&j=3,\cdots,l\\
\frac{2}{a^{2}}f(l^{2j-3})+\frac{2}{a^{2}}f(l^{2j-4})+2\cdot2^{2(j-1)}+2^{2(j-1)}&j=l+1
\end{array}
\right.
\end{equation*}
and in $t_{l}$,
\begin{equation*}
f((l+1)^{2j-2})=\left\{
\begin{array}{ll}
\frac{2}{a^{2}}(2l+1)+\frac{2}{a^{2}}+2l(l+1)&j=2\\
\frac{2}{a^{2}}f(l^{2j-3})+\frac{2}{a^{2}}f(l^{2j-4})+f(l^{2j-2})&j=3,\cdots,l\\
\frac{2}{a^{2}}f(l^{2j-3})+\frac{2}{a^{2}}f(l^{2j-4})+2^{2(j-1)}&j=l+1
\end{array}
\right.
\end{equation*}

\begin{enumerate}[I.]
\item Prove Eq. (33)
\end{enumerate}
When $j=2$
$$
\frac{f((l+1)^5)}{f((l+1)^3)}=\frac{\frac{2}{a^{2}}f(l^3)+\frac{2}{a^{2}}f(l^2)+2f(l^4)+f(l^5)}{\frac{2}{a^{2}}(2l+1)+\frac{2}{a^{2}}+2\cdot2l(l+1)+\frac{2}{3}l(l+1)(2l+1)}
$$
Due to
$$
f(l^4)<a^{2}f(l^2)\ \text{ and }\ f(l^5)<a^{2}f(l^3),
$$
and
$$
f(l^2)=2l(l+1)\ \text{ and }\ f(l^3)=\frac{2}{3}l(l+1)(2l+1),
$$
then
\begin{equation*}
\begin{split}
\frac{f((l+1)^5)}{f((l+1)^3)}&<\frac{\frac{2}{a^{2}}f(l^3)+\frac{2}{a^{2}}f(l^2)+2a^{2}f(l^2)+a^{2}f(l^3)}{\frac{2}{a^{2}}(2l+1)+\frac{2}{a^{2}}+2\cdot2l(l+1)+\frac{2}{3}l(l+1)(2l+1)}\\
&=a^{2}\frac{\frac{1}{a^{2}}(4l^{3}+12l^{2}+8l)+a^{2}(2l^{3}+9l^{2}+7l)}{(6l+6)+a^{2}(2l^{3}+9l^{2}+7l)}\\
&\leq
a^{2}\frac{\frac{1}{l^{2}}(4l^{3}+12l^{2}+8l)+a^{2}(2l^{3}+9l^{2}+7l)}{(6l+6)+a^{2}(2l^{3}+9l^{2}+7l)}\\
&=a^{2}\frac{(4l+12+\frac{8}{l})+a^{2}(2l^{3}+9l^{2}+7l)}{(6l+6)+a^{2}(2l^{3}+9l^{2}+7l)}
\end{split}
\end{equation*}
Because $l\geq4$,
$$
4l+12+\frac{8}{l}\leq6l+6.
$$
Hence,
\begin{equation}
\frac{f((l+1)^5)}{f((l+1)^3)}<a^{2}.
\end{equation}

When $j=3,\cdots,(l+1)-2$
\begin{equation}
\begin{split}
\frac{f((l+1)^{2j+1})}{f((l+1)^{2j-1})}&=\frac{\frac{2}{a^{2}}f(l^{2j-1})+\frac{2}{a^{2}}f(l^{2j-2})+2f(l^{2j})+f(l^{2j+1})}{\frac{2}{a^{2}}f(l^{2j-3})+\frac{2}{a^{2}}f(l^{2j-4})+2f(l^{2j-2})+f(l^{2j-1})}\\
&<a^{2}\frac{\frac{2}{a^{4}}f(l^{2j-1})+\frac{2}{a^{4}}f(l^{2j-2})+2f(l^{2j-2})+f(l^{2j-1})}{\frac{2}{a^{4}}f(l^{2j-1})+\frac{2}{a^{4}}f(l^{2j-2})+2f(l^{2j-2})+f(l^{2j-1})}\\
&=a^{2}
\end{split}
\end{equation}

When $j=(l+1)-1=l$
\begin{equation}
\begin{split}
\frac{f((l+1)^{2j+1})}{f((l+1)^{2j-1})}&=\frac{\frac{2}{a^{2}}f(l^{2j-1})+\frac{2}{a^{2}}f(l^{2j-2})+2\cdot2^{2j}+2^{2j}}{\frac{2}{a^{2}}f(l^{2j-3})+\frac{2}{a^{2}}f(l^{2j-4})+2f(l^{2j-2})+f(l^{2j-1})}\\
&<a^{2}\frac{\frac{2}{a^{4}}f(l^{2j-1})+\frac{2}{a^{4}}f(l^{2j-2})+2f(l^{2j-2})+f(l^{2j-1})}{\frac{2}{a^{4}}f(l^{2j-1})+\frac{2}{a^{4}}f(l^{2j-2})+2f(l^{2j-2})+f(l^{2j-1})}\\
&=a^{2}
\end{split}
\end{equation}

According to Eq. (35)-(37), and mathematical induction, Eq. (33) is
proved.

\begin{enumerate}[II.]
\item Prove Eq. (34)
\end{enumerate}
When $j=2$
$$
\frac{f((l+1)^4)}{f((l+1)^2)}=\frac{\frac{2}{a^{2}}f(l^3)+\frac{2}{a^{2}}f(l^2)+f(l^4)}{\frac{2}{a^{2}}(2l+1)+\frac{2}{a^{2}}+2l(l+1)}
$$
Due to
$$
f(l^4)<a^{2}f(l^2),
$$
and
$$
f(l^2)=2l(l+1)\ \text{ and }\ f(l^3)=\frac{2}{3}l(l+1)(2l+1),
$$
then
\begin{equation*}
\begin{split}
\frac{f((l+1)^4)}{f((l+1)^2)}&<\frac{\frac{2}{a^{2}}f(l^3)+\frac{2}{a^{2}}f(l^2)+a^{2}f(l^2)}{\frac{2}{a^{2}}(2l+1)+\frac{2}{a^{2}}+2l(l+1)}\\
&=a^{2}\frac{\frac{1}{a^{2}}(4l^{3}+12l^{2}+8l)+3a^{2}l(l+1)}{(6l+6)+3a^{2}l(l+1)}\\
&\leq
a^{2}\frac{\frac{1}{l^{2}}(4l^{3}+12l^{2}+8l)+3a^{2}l(l+1)}{(6l+6)+3a^{2}l(l+1)}\\
&=a^{2}\frac{(4l+12+\frac{8}{l})+3a^{2}l(l+1)}{(6l+6)+3a^{2}l(l+1)}
\end{split}
\end{equation*}
Because $l\geq4$,
$$
4l+12+\frac{8}{l}\leq6l+6.
$$
Hence,
\begin{equation}
\frac{f((l+1)^4)}{f((l+1)^2)}<a^{2}.
\end{equation}

When $j=3,\cdots,(l+1)-2$
\begin{equation}
\begin{split}
\frac{f((l+1)^{2j})}{f((l+1)^{2j-2})}&=\frac{\frac{2}{a^{2}}f(l^{2j-1})+\frac{2}{a^{2}}f(l^{2j-2})+f(l^{2j})}{\frac{2}{a^{2}}f(l^{2j-3})+\frac{2}{a^{2}}f(l^{2j-4})+f(l^{2j-2})}\\
&<a^{2}\frac{\frac{2}{a^{4}}f(l^{2j-1})+\frac{2}{a^{4}}f(l^{2j-2})+f(l^{2j-2})}{\frac{2}{a^{4}}f(l^{2j-1})+\frac{2}{a^{4}}f(l^{2j-2})+f(l^{2j-2})}\\
&=a^{2}
\end{split}
\end{equation}

When $j=(l+1)-1=l$
\begin{equation}
\begin{split}
\frac{f((l+1)^{2j})}{f((l+1)^{2j-2})}&=\frac{\frac{2}{a^{2}}f(l^{2j-1})+\frac{2}{a^{2}}f(l^{2j-2})+2^{2j}}{\frac{2}{a^{2}}f(l^{2j-3})+\frac{2}{a^{2}}f(l^{2j-4})+f(l^{2j-2})}\\
&<a^{2}\frac{\frac{2}{a^{4}}f(l^{2j-1})+\frac{2}{a^{4}}f(l^{2j-2})+f(l^{2j-2})}{\frac{2}{a^{4}}f(l^{2j-1})+\frac{2}{a^{4}}f(l^{2j-2})+f(l^{2j-2})}\\
&=a^{2}
\end{split}
\end{equation}
According to Eq. (38)-(40), and mathematical induction, Eq. (34) is
proved.

Synthesize I and II, Lemma 3 is proved.

\end{proof}

\begin{theorem}
When the unitary operations Step 3.1 and 3.2 are repeated
\begin{equation}
\hat{i}=\left\lceil-1+\frac{1}{2}\sqrt{4-\frac{2}{3}c+A+B}-\frac{1}{2}\sqrt{8-\frac{4}{3}c-A-B}\right\rceil
\end{equation}
times, the probability of $|k_{A_{0}}\rangle$ being measured is the
biggest, where
\begin{equation*}
\begin{split}
&b=4,\ \ c=2-3a^{2},\ \
d=-1-6a^{2},\ \ e=\frac{3}{2}a^{4}-\frac{3}{2}a^{2}\\
&\alpha=c^{2}-3bd+12e,\ \ \beta=2c^{3}-9bcd+27d^{2}+27b^{2}e-72ce\\
&A=\frac{\sqrt[3]{2}\alpha}{3\sqrt[3]{\beta+\sqrt{-4\alpha^{3}+\beta^{2}}}},\
\
B=\frac{\sqrt[3]{\beta+\sqrt{-4\alpha^{3}+\beta^{2}}}}{3\sqrt[3]{2}}
\end{split}
\end{equation*}
\end{theorem}
\begin{proof}
According to the periodicity of Grover's algorithm, if $i$ is the
smallest one that makes $t_{(i+1)0}<t_{i0}$, $t_{i0}^{2}$ is the
biggest probability of $k_{A_{0}}$ being measured. According to Eq.
(26)
\begin{equation*}
t_{(i+1)0}=-\frac{2t_{i0}}{a^{2}}-\frac{2t_{i}}{a^{2}}+2t_{i}+t_{i0}<t_{i0}
\end{equation*}
i.e.
\begin{equation*}
a^{2}t_{i}-t_{i}-t_{i0}<0
\end{equation*}
Due to $t_{i}\geq0$, if
\begin{equation}
a^{2}t_{i}-t_{i0}<0,
\end{equation}
the theorem is proved.

Substitute Eq. (29) and (30) into it
\begin{equation}
\begin{split}
&a^{2}t_{i}-t_{i0}\\
=&a-\frac{2i^{2}+4i+1}{a}+\frac{\frac{2}{3}i^{4}+\frac{8}{3}i^{3}+\frac{4}{3}i^{2}-\frac{2}{3}i}{a^{3}}\\
&+\sum_{j=4}^{i}(-1)^{j-1}\frac{f(i^{2j-2})}{a^{2j-3}}+\sum_{j=3}^{i}(-1)^{j}\frac{f(i^{2j-1})}{a^{2j-1}}+(-1)^{i}\frac{2^{2i}}{a^{2i-1}}+(-1)^{i+1}\frac{2^{2i}}{a^{2i+1}}
\end{split}
\end{equation}

When $i$ is an even number,
\begin{equation*}
\begin{split}
&a^{2}t_{i}-t_{i0}\\
=&a-\frac{2i^{2}+4i+1}{a}+\frac{\frac{2}{3}i^{4}+\frac{8}{3}i^{3}+\frac{4}{3}i^{2}-\frac{2}{3}i}{a^{3}}\\
&+\sum_{j=3,j=j+2}^{i-3}\left(-\frac{f(i^{2j-1})+f(i^{2j})}{a^{2j-1}}+\frac{f(i^{2j+1})+f(i^{2j+2})}{a^{2j+1}}\right)\\
&-\frac{f(i^{2i-3})+f(i^{2i-2})}{a^{2i-3}}+\frac{f(i^{2i-1})}{a^{2i-1}}+\frac{2^{2i}}{a^{2i-1}}-\frac{2^{2i}}{a^{2i+1}}\\
=&a-\frac{2i^{2}+4i+1}{a}+\frac{\frac{2}{3}i^{4}+\frac{8}{3}i^{3}+\frac{4}{3}i^{2}-\frac{2}{3}i}{a^{3}}\\
&+\sum_{j=3,j=j+2}^{i-3}\left(\frac{f(i^{2j+1})-a^{2}f(i^{2j-1})+f(i^{2j+2})-a^{2}f(i^{2j})}{a^{2j+1}}\right)\\
&+\frac{f(i^{2i-1})-a^{2}f(i^{2i-3})+2^{2i}-a^{2}f(i^{2i-2})}{a^{2i-1}}-\frac{2^{2i}}{a^{2i+1}}\\
\end{split}
\end{equation*}
According to Lemma 3 and 4,
\begin{equation}
a^{2}t_{i}-t_{i0}<a-\frac{2i^{2}+4i+1}{a}+\frac{\frac{2}{3}i^{4}+\frac{8}{3}i^{3}+\frac{4}{3}i^{2}-\frac{2}{3}i}{a^{3}}
\end{equation}

When $i$ is an odd number,
\begin{equation*}
\begin{split}
&a^{2}t_{i}-t_{i0}\\
=&a-\frac{2i^{2}+4i+1}{a}+\frac{\frac{2}{3}i^{4}+\frac{8}{3}i^{3}+\frac{4}{3}i^{2}-\frac{2}{3}i}{a^{3}}\\
&+\sum_{j=3,j=j+2}^{i-2}\left(-\frac{f(i^{2j-1})+f(i^{2j})}{a^{2j-1}}+\frac{f(i^{2j+1})+f(i^{2j+2})}{a^{2j+1}}\right)\\
&-\frac{f(i^{2i-1})}{a^{2i-1}}-\frac{2^{2i}}{a^{2i-1}}+\frac{2^{2i}}{a^{2i+1}}\\
=&a-\frac{2i^{2}+4i+1}{a}+\frac{\frac{2}{3}i^{4}+\frac{8}{3}i^{3}+\frac{4}{3}i^{2}-\frac{2}{3}i}{a^{3}}\\
&+\sum_{j=3,j=j+2}^{i-2}\left(\frac{f(i^{2j+1})-a^{2}f(i^{2j-1})+f(i^{2j+2})-a^{2}f(i^{2j})}{a^{2j+1}}\right)\\
&-\frac{f(i^{2i-1})}{a^{2i-1}}-\frac{2^{2i}}{a^{2i+1}}(a^{2}-1)
\end{split}
\end{equation*}
According to Lemma 3 and 4,
\begin{equation}
a^{2}t_{i}-t_{i0}<a-\frac{2i^{2}+4i+1}{a}+\frac{\frac{2}{3}i^{4}+\frac{8}{3}i^{3}+\frac{4}{3}i^{2}-\frac{2}{3}i}{a^{3}}
\end{equation}

Eq. (44) and (45) indicate that $a^{2}t_{i}-t_{i0}$ is
less than
$a-\frac{2i^{2}+4i+1}{a}+\frac{\frac{2}{3}i^{4}+\frac{8}{3}i^{3}+\frac{4}{3}i^{2}-\frac{2}{3}i}{a^{3}}$.
If the latter is less than 0, Eq. (42) will be satisfied, i.e., the
$i$ that makes
$a-\frac{2i^{2}+4i+1}{a}+\frac{\frac{2}{3}i^{4}+\frac{8}{3}i^{3}+\frac{4}{3}i^{2}-\frac{2}{3}i}{a^{3}}$
less than 0 is the suitable number of iterations.
\begin{equation*}
\begin{split}
&a-\frac{2i^{2}+4i+1}{a}+\frac{\frac{2}{3}i^{4}+\frac{8}{3}i^{3}+\frac{4}{3}i^{2}-\frac{2}{3}i}{a^{3}}\\
=&\frac{2}{3a^{3}}\left(i^{4}+4i^{3}+(2-3a^{2})i^{2}+(-1-6a^{2})i+\frac{3}{2}a^{4}-\frac{3}{2}a^{2}\right)<0
\end{split}
\end{equation*}
i.e.,
\begin{equation}
i^{4}+4i^{3}+(2-3a^{2})i^{2}+(-1-6a^{2})i+\frac{3}{2}a^{4}-\frac{3}{2}a^{2}<0
\end{equation}
It is a biqudratic inequality, where $b=4$, $c=2-3a^{2}$,
$d=-1-6a^{2}$ and $e=\frac{3}{2}a^{4}-\frac{3}{2}a^{2}$. Solving
this inequality, then
\begin{equation}
\begin{split}
&-1+\frac{1}{2}\sqrt{4-\frac{2}{3}c+A+B}-\frac{1}{2}\sqrt{8-\frac{4}{3}c-A-B}\\
<i<&-1+\frac{1}{2}\sqrt{4-\frac{2}{3}c+A+B}+\frac{1}{2}\sqrt{8-\frac{4}{3}c-A-B}
\end{split}
\end{equation}
where
\begin{equation*}
\alpha=c^{2}-3bd+12e
\end{equation*}
\begin{equation*}
\beta=2c^{3}-9bcd+27d^{2}+27b^{2}e-72ce
\end{equation*}
\begin{equation*}
A=\frac{\sqrt[3]{2}\alpha}{3\sqrt[3]{\beta+\sqrt{-4\alpha^{3}+\beta^{2}}}}
\end{equation*}
\begin{equation*}
B=\frac{\sqrt[3]{\beta+\sqrt{-4\alpha^{3}+\beta^{2}}}}{3\sqrt[3]{2}}
\end{equation*}

Hence,
\begin{equation*}
i=\left\lceil-1+\frac{1}{2}\sqrt{4-\frac{2}{3}c+A+B}-\frac{1}{2}\sqrt{8-\frac{4}{3}c-A-B}\right\rceil
\end{equation*}
We represent the $i$ at this time as $\hat{i}$.
\end{proof}

\begin{theorem}
\begin{equation}
\hat{i}\approx0.7962a-0.6057
\end{equation}
\end{theorem}
\begin{proof}
According to Theorem 1, we can get Table 1.
\begin{table}
\centering \caption{$\hat{i}$ is changing with $a$ according to Eq.
(41).}
\begin{tabular}{|c|c|c|c|c|c|}
\hline
$a$ & 4 &8&16&32&64 \\
\hline
$\hat{i}$ &3&6&12&25&50 \\
\hline
$a$ & 128&256&512&1024&2048 \\
\hline
$\hat{i}$ &101&203&407&815&1630 \\
\hline
$a$ & 4096&8192&16384&32768&65536 \\
\hline
$\hat{i}$ &3261&6522&13044&26090&52180 \\
\hline
\end{tabular}
\end{table}
Table 1 covers the common image sizes: from $4\times4$ to
$65536\times65536$. According to Table 1 and by using polynomial
curve fitting, Theorem 2 is proved.
\end{proof}

Seemingly, $t_{\hat{i}0}$ can be gained only by substituting Eq.
(41) or (48) into Eq. (29). However, these equations are too
complicated to calculate. Therefore, we simplify them to give the
lower bound of $t_{\hat{i}0}$.

\begin{theorem}
The lower bound of the probability of $|k_{A_{0}}\rangle$ being
measured is
\begin{equation}
(0.9194+0.0567a^{-1}+0.2302a^{-2}-0.0336a^{-3})^{2}
\end{equation}
\end{theorem}
\begin{proof}
\begin{equation*}
\begin{split}
t_{i0}=&\frac{2i+1}{a}-\frac{\frac{2}{3}i(i+1)(2i+1)}{a^{3}}+\sum_{j=3}^{i}(-1)^{j-1}\frac{f(i^{2j-1})}{a^{2j-1}}+(-1)^{i}\frac{2^{2i}}{a^{2i+1}}\\
=&\frac{2i+1}{a}-\frac{\frac{2}{3}i(i+1)(2i+1)}{a^{3}}+\sum_{j=3,j=j+2}^{i}\left(\frac{f(i^{2j-1})}{a^{2j-1}}-\frac{f(i^{2j+1})}{a^{2j+1}}\right)+(-1)^{i}\frac{2^{2i}}{a^{2i+1}}
\end{split}
\end{equation*}
Similar with the proof of Theorem 1,
\begin{equation*}
t_{i0}>\frac{2i+1}{a}-\frac{\frac{2}{3}i(i+1)(2i+1)}{a^{3}}
\end{equation*}
Substitute Eq. (48) into it
\begin{equation*}
t_{\hat{i}0}>0.9194+0.0567a^{-1}+0.2302a^{-2}-0.0336a^{-3}
\end{equation*}
Hence, the theorem is proved.
\end{proof}

\subsection{An example}
Fig. 7 gives an example in which image $B$ is in the middle of image
$A$ and they are extracted from the real gray-scale image ``Lena''.
\begin{figure}[h]
  \centering
  \includegraphics[width=10cm,keepaspectratio]{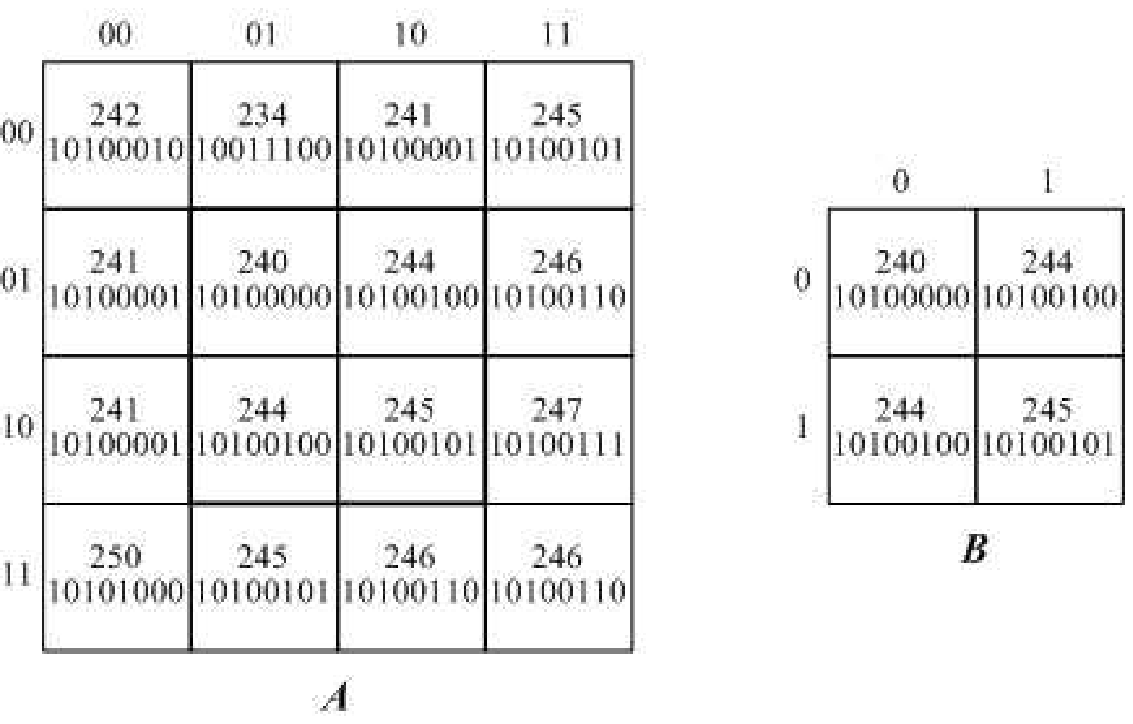}
  \caption{An example.}
  \label{fig:7}
\end{figure}

Initially,
\begin{tiny}
\begin{equation*}
\begin{split}
\Psi_{0}&=|g\rangle\otimes|f\rangle\otimes|A\rangle\otimes|B\rangle\\
&=|g\rangle\otimes\left(\frac{1}{8}|0\rangle|10100010\rangle|0000\rangle|10100000\rangle|00\rangle+\frac{1}{8}|0\rangle|10100010\rangle|0000\rangle|10100100\rangle|01\rangle\right.\\
&+\frac{1}{8}|0\rangle|10100010\rangle|0000\rangle|10100100\rangle|10\rangle+\frac{1}{8}|0\rangle|10100010\rangle|0000\rangle|10100101\rangle|11\rangle+\frac{1}{8}|0\rangle|10011100\rangle|0001\rangle|10100000\rangle|00\rangle\\
&+\frac{1}{8}|0\rangle|10011100\rangle|0001\rangle|10100100\rangle|01\rangle+\frac{1}{8}|0\rangle|10011100\rangle|0001\rangle|10100100\rangle|10\rangle+\frac{1}{8}|0\rangle|10011100\rangle|0001\rangle|10100101\rangle|11\rangle\\
&+\frac{1}{8}|0\rangle|10100001\rangle|0010\rangle|10100000\rangle|00\rangle+\frac{1}{8}|0\rangle|10100001\rangle|0010\rangle|10100100\rangle|01\rangle+\frac{1}{8}|0\rangle|10100001\rangle|0010\rangle|10100100\rangle|10\rangle\\
&+\frac{1}{8}|0\rangle|10100001\rangle|0010\rangle|10100101\rangle|11\rangle+\frac{1}{8}|0\rangle|10100101\rangle|0011\rangle|10100000\rangle|00\rangle+\frac{1}{8}|0\rangle|10100101\rangle|0011\rangle|10100100\rangle|01\rangle\\
&+\frac{1}{8}|0\rangle|10100101\rangle|0011\rangle|10100100\rangle|10\rangle+\frac{1}{8}|0\rangle|10100101\rangle|0011\rangle|10100101\rangle|11\rangle+\frac{1}{8}|0\rangle|10100001\rangle|0100\rangle|10100000\rangle|00\rangle\\
&+\frac{1}{8}|0\rangle|10100001\rangle|0100\rangle|10100100\rangle|01\rangle+\frac{1}{8}|0\rangle|10100001\rangle|0100\rangle|10100100\rangle|10\rangle+\frac{1}{8}|0\rangle|10100001\rangle|0100\rangle|10100101\rangle|11\rangle\\
&+\frac{1}{8}|0\rangle|10100000\rangle|0101\rangle|10100000\rangle|00\rangle+\frac{1}{8}|0\rangle|10100000\rangle|0101\rangle|10100100\rangle|01\rangle+\frac{1}{8}|0\rangle|10100000\rangle|0101\rangle|10100100\rangle|10\rangle\\
&+\frac{1}{8}|0\rangle|10100000\rangle|0101\rangle|10100101\rangle|11\rangle+\frac{1}{8}|0\rangle|10100100\rangle|0110\rangle|10100000\rangle|00\rangle+\frac{1}{8}|0\rangle|10100100\rangle|0110\rangle|10100100\rangle|01\rangle\\
&+\frac{1}{8}|0\rangle|10100100\rangle|0110\rangle|10100100\rangle|10\rangle+\frac{1}{8}|0\rangle|10100100\rangle|0110\rangle|10100101\rangle|11\rangle+\frac{1}{8}|0\rangle|10100110\rangle|0111\rangle|10100000\rangle|00\rangle\\
&+\frac{1}{8}|0\rangle|10100110\rangle|0111\rangle|10100100\rangle|01\rangle+\frac{1}{8}|0\rangle|10100110\rangle|0111\rangle|10100100\rangle|10\rangle+\frac{1}{8}|0\rangle|10100110\rangle|0111\rangle|10100101\rangle|11\rangle\\
&+\frac{1}{8}|0\rangle|10100001\rangle|1000\rangle|10100000\rangle|00\rangle+\frac{1}{8}|0\rangle|10100001\rangle|1000\rangle|10100100\rangle|01\rangle+\frac{1}{8}|0\rangle|10100001\rangle|1000\rangle|10100100\rangle|10\rangle\\
&+\frac{1}{8}|0\rangle|10100001\rangle|1000\rangle|10100101\rangle|11\rangle+\frac{1}{8}|0\rangle|10100100\rangle|1001\rangle|10100000\rangle|00\rangle+\frac{1}{8}|0\rangle|10100100\rangle|1001\rangle|10100100\rangle|01\rangle\\
&+\frac{1}{8}|0\rangle|10100100\rangle|1001\rangle|10100100\rangle|10\rangle+\frac{1}{8}|0\rangle|10100100\rangle|1001\rangle|10100101\rangle|11\rangle+\frac{1}{8}|0\rangle|10100101\rangle|1010\rangle|10100000\rangle|00\rangle\\
&+\frac{1}{8}|0\rangle|10100101\rangle|1010\rangle|10100100\rangle|01\rangle+\frac{1}{8}|0\rangle|10100101\rangle|1010\rangle|10100100\rangle|10\rangle+\frac{1}{8}|0\rangle|10100101\rangle|1010\rangle|10100101\rangle|11\rangle\\
&+\frac{1}{8}|0\rangle|10100111\rangle|1011\rangle|10100000\rangle|00\rangle+\frac{1}{8}|0\rangle|10100111\rangle|1011\rangle|10100100\rangle|01\rangle+\frac{1}{8}|0\rangle|10100111\rangle|1011\rangle|10100100\rangle|10\rangle\\
&+\frac{1}{8}|0\rangle|10100111\rangle|1011\rangle|10100101\rangle|11\rangle+\frac{1}{8}|0\rangle|10101000\rangle|1100\rangle|10100000\rangle|00\rangle+\frac{1}{8}|0\rangle|10101000\rangle|1100\rangle|10100100\rangle|01\rangle\\
&+\frac{1}{8}|0\rangle|10101000\rangle|1100\rangle|10100100\rangle|10\rangle+\frac{1}{8}|0\rangle|10101000\rangle|1100\rangle|10100101\rangle|11\rangle+\frac{1}{8}|0\rangle|10100101\rangle|1101\rangle|10100000\rangle|00\rangle\\
&+\frac{1}{8}|0\rangle|10100101\rangle|1101\rangle|10100100\rangle|01\rangle+\frac{1}{8}|0\rangle|10100101\rangle|1101\rangle|10100100\rangle|10\rangle+\frac{1}{8}|0\rangle|10100101\rangle|1101\rangle|10100101\rangle|11\rangle\\
&+\frac{1}{8}|0\rangle|10100110\rangle|1110\rangle|10100000\rangle|00\rangle+\frac{1}{8}|0\rangle|10100110\rangle|1110\rangle|10100100\rangle|01\rangle+\frac{1}{8}|0\rangle|10100110\rangle|1110\rangle|10100100\rangle|10\rangle\\
&+\frac{1}{8}|0\rangle|10100110\rangle|1110\rangle|10100101\rangle|11\rangle+\frac{1}{8}|0\rangle|10100110\rangle|1111\rangle|10100000\rangle|00\rangle+\frac{1}{8}|0\rangle|10100110\rangle|1111\rangle|10100100\rangle|01\rangle\\
&\left.+\frac{1}{8}|0\rangle|10100110\rangle|1111\rangle|10100100\rangle|10\rangle+\frac{1}{8}|0\rangle|10100110\rangle|1111\rangle|10100101\rangle|11\rangle\right)
\end{split}
\end{equation*}
\end{tiny}

Step 1 changes $I_{A}^{i}$ to $I_{A}^{i}\oplus I_{B}^{i}$ and Step 2
changes $|f\rangle$ to $|1\rangle$ when
$|I_{A}\rangle=|0\rangle^{\otimes q}$ and
$|k_{B}\rangle=|0\rangle^{\otimes m}$. Hence,
\begin{tiny}
\begin{equation*}
\begin{split}
\Psi_{2}&=|g\rangle\otimes|f\rangle\otimes|A\rangle\otimes|B\rangle\\
&=|g\rangle\otimes\left(\frac{1}{8}|0\rangle|00000010\rangle|0000\rangle|10100000\rangle|00\rangle+\frac{1}{8}|0\rangle|00000110\rangle|0000\rangle|10100100\rangle|01\rangle\right.\\
&+\frac{1}{8}|0\rangle|00000110\rangle|0000\rangle|10100100\rangle|10\rangle+\frac{1}{8}|0\rangle|00000111\rangle|0000\rangle|10100101\rangle|11\rangle+\frac{1}{8}|0\rangle|00111100\rangle|0001\rangle|10100000\rangle|00\rangle\\
&+\frac{1}{8}|0\rangle|00111000\rangle|0001\rangle|10100100\rangle|01\rangle+\frac{1}{8}|0\rangle|00111000\rangle|0001\rangle|10100100\rangle|10\rangle+\frac{1}{8}|0\rangle|00111001\rangle|0001\rangle|10100101\rangle|11\rangle\\
&+\frac{1}{8}|0\rangle|00000001\rangle|0010\rangle|10100000\rangle|00\rangle+\frac{1}{8}|0\rangle|00000101\rangle|0010\rangle|10100100\rangle|01\rangle+\frac{1}{8}|0\rangle|00000101\rangle|0010\rangle|10100100\rangle|10\rangle\\
&+\frac{1}{8}|0\rangle|00000100\rangle|0010\rangle|10100101\rangle|11\rangle+\frac{1}{8}|0\rangle|00000101\rangle|0011\rangle|10100000\rangle|00\rangle+\frac{1}{8}|0\rangle|00000001\rangle|0011\rangle|10100100\rangle|01\rangle\\
&+\frac{1}{8}|0\rangle|00000001\rangle|0011\rangle|10100100\rangle|10\rangle+\frac{1}{8}|0\rangle|00000000\rangle|0011\rangle|10100101\rangle|11\rangle+\frac{1}{8}|0\rangle|00000001\rangle|0100\rangle|10100000\rangle|00\rangle\\
&+\frac{1}{8}|0\rangle|00000101\rangle|0100\rangle|10100100\rangle|01\rangle+\frac{1}{8}|0\rangle|00000101\rangle|0100\rangle|10100100\rangle|10\rangle+\frac{1}{8}|0\rangle|00000100\rangle|0100\rangle|10100101\rangle|11\rangle\\
&+\boxed{\frac{1}{8}|1\rangle|00000000\rangle|0101\rangle|10100000\rangle|00\rangle}+\frac{1}{8}|0\rangle|00000100\rangle|0101\rangle|10100100\rangle|01\rangle+\frac{1}{8}|0\rangle|00000100\rangle|0101\rangle|10100100\rangle|10\rangle\\
&+\frac{1}{8}|0\rangle|00000101\rangle|0101\rangle|10100101\rangle|11\rangle+\frac{1}{8}|0\rangle|00000100\rangle|0110\rangle|10100000\rangle|00\rangle+\frac{1}{8}|0\rangle|00000000\rangle|0110\rangle|10100100\rangle|01\rangle\\
&+\frac{1}{8}|0\rangle|00000000\rangle|0110\rangle|10100100\rangle|10\rangle+\frac{1}{8}|0\rangle|00000001\rangle|0110\rangle|10100101\rangle|11\rangle+\frac{1}{8}|0\rangle|00000110\rangle|0111\rangle|10100000\rangle|00\rangle\\
&+\frac{1}{8}|0\rangle|00000010\rangle|0111\rangle|10100100\rangle|01\rangle+\frac{1}{8}|0\rangle|00000010\rangle|0111\rangle|10100100\rangle|10\rangle+\frac{1}{8}|0\rangle|00000011\rangle|0111\rangle|10100101\rangle|11\rangle\\
&+\frac{1}{8}|0\rangle|00000001\rangle|1000\rangle|10100000\rangle|00\rangle+\frac{1}{8}|0\rangle|00000101\rangle|1000\rangle|10100100\rangle|01\rangle+\frac{1}{8}|0\rangle|00000101\rangle|1000\rangle|10100100\rangle|10\rangle\\
&+\frac{1}{8}|0\rangle|00000100\rangle|1000\rangle|10100101\rangle|11\rangle+\frac{1}{8}|0\rangle|00000100\rangle|1001\rangle|10100000\rangle|00\rangle+\frac{1}{8}|0\rangle|00000000\rangle|1001\rangle|10100100\rangle|01\rangle\\
&+\frac{1}{8}|0\rangle|00000000\rangle|1001\rangle|10100100\rangle|10\rangle+\frac{1}{8}|0\rangle|00000001\rangle|1001\rangle|10100101\rangle|11\rangle+\frac{1}{8}|0\rangle|00000101\rangle|1010\rangle|10100000\rangle|00\rangle\\
&+\frac{1}{8}|0\rangle|00000001\rangle|1010\rangle|10100100\rangle|01\rangle+\frac{1}{8}|0\rangle|00000001\rangle|1010\rangle|10100100\rangle|10\rangle+\frac{1}{8}|0\rangle|00000000\rangle|1010\rangle|10100101\rangle|11\rangle\\
&+\frac{1}{8}|0\rangle|00000111\rangle|1011\rangle|10100000\rangle|00\rangle+\frac{1}{8}|0\rangle|00000011\rangle|1011\rangle|10100100\rangle|01\rangle+\frac{1}{8}|0\rangle|00000011\rangle|1011\rangle|10100100\rangle|10\rangle\\
&+\frac{1}{8}|0\rangle|00000010\rangle|1011\rangle|10100101\rangle|11\rangle+\frac{1}{8}|0\rangle|00001000\rangle|1100\rangle|10100000\rangle|00\rangle+\frac{1}{8}|0\rangle|00001100\rangle|1100\rangle|10100100\rangle|01\rangle\\
&+\frac{1}{8}|0\rangle|00001100\rangle|1100\rangle|10100100\rangle|10\rangle+\frac{1}{8}|0\rangle|00001101\rangle|1100\rangle|10100101\rangle|11\rangle+\frac{1}{8}|0\rangle|00000101\rangle|1101\rangle|10100000\rangle|00\rangle\\
&+\frac{1}{8}|0\rangle|00000001\rangle|1101\rangle|10100100\rangle|01\rangle+\frac{1}{8}|0\rangle|00000001\rangle|1101\rangle|10100100\rangle|10\rangle+\frac{1}{8}|0\rangle|00000000\rangle|1101\rangle|10100101\rangle|11\rangle\\
&+\frac{1}{8}|0\rangle|00000110\rangle|1110\rangle|10100000\rangle|00\rangle+\frac{1}{8}|0\rangle|00000010\rangle|1110\rangle|10100100\rangle|01\rangle+\frac{1}{8}|0\rangle|00000010\rangle|1110\rangle|10100100\rangle|10\rangle\\
&+\frac{1}{8}|0\rangle|00000011\rangle|1110\rangle|10100101\rangle|11\rangle+\frac{1}{8}|0\rangle|00000110\rangle|1111\rangle|10100000\rangle|00\rangle+\frac{1}{8}|0\rangle|00000010\rangle|1111\rangle|10100100\rangle|01\rangle\\
&\left.+\frac{1}{8}|0\rangle|00000010\rangle|1111\rangle|10100100\rangle|10\rangle+\frac{1}{8}|0\rangle|00000011\rangle|1111\rangle|10100101\rangle|11\rangle\right)
\end{split}
\end{equation*}
\end{tiny}
The boxed pixel is the one that we want to find out.

Hence, in the subspace $|k_{A}\rangle$.
\begin{equation*}
\Psi_{30}=\left(\frac{1}{4},\frac{1}{4},\frac{1}{4},\frac{1}{4},\frac{1}{4},\boxed{\frac{1}{4}},\frac{1}{4},\frac{1}{4},\frac{1}{4},\frac{1}{4},\frac{1}{4},\frac{1}{4},\frac{1}{4},\frac{1}{4},\frac{1}{4},\frac{1}{4}\right)^{\text{T}}
\end{equation*}
It should be iterated
$$
\hat{i}=\left\lceil-1+\frac{1}{2}\sqrt{4-\frac{2}{3}c+A+B}-\frac{1}{2}\sqrt{8-\frac{4}{3}c-A-B}\right\rceil=3
$$
times.
\begin{enumerate}[$\bullet$]

\item for the first iteration:
\begin{equation*}
\begin{split}
&\overline{t}_{1}=\left. \left(15\cdot\frac{1}{4}-\frac{1}{4}\right)
\middle / 16 \right.=\frac{7}{32}\\
&t_{10}=2\cdot\frac{7}{32}+\frac{1}{4}=\frac{11}{16},\ \ \
t_{1}=2\cdot\frac{7}{32}-\frac{1}{4}=\frac{3}{16}
\end{split}
\end{equation*}

\item for the second iteration:
\begin{equation*}
\begin{split}
&\overline{t}_{2}=\left.
\left(15\cdot\frac{3}{16}-\frac{11}{16}\right)
\middle / 16 \right.=\frac{17}{128}\\
&t_{20}=2\cdot\frac{17}{128}+\frac{11}{16}=\frac{61}{64},\ \ \
t_{2}=2\cdot\frac{17}{128}-\frac{3}{16}=\frac{5}{64}
\end{split}
\end{equation*}

\item for the third iteration:
\begin{equation*}
\begin{split}
&\overline{t}_{3}=\left.
\left(15\cdot\frac{5}{64}-\frac{61}{64}\right)
\middle / 16 \right.=\frac{7}{512}\\
&t_{30}=2\cdot\frac{7}{512}+\frac{61}{64}=\frac{251}{256},\ \ \
t_{3}=2\cdot\frac{7}{512}-\frac{5}{64}=-\frac{13}{256}
\end{split}
\end{equation*}
\end{enumerate}

After 3 times of iterations, the sixth basis $|0101\rangle$ will be
measured with probability $\left(\frac{251}{256}\right)^{2}=0.9613$
which is much higher than the probability
$\left(-\frac{13}{256}\right)^{2}=0.002579$ of other pixels being
measured.

In this example, if it is iterated 4 times, $t_{40}$ will be
$\frac{781}{1024}<\frac{251}{256}$. Hence, Theorem 1 and 2 are
verified.

Moreover,
$t_{30}^{2}=0.9613>(0.9194+0.0567a^{-1}+0.2302a^{-2}-0.0336a^{-3})^{2}=0.8976$.
Hence, Theorem 3 is verified.

\section{Complexity analysis}

In this scheme, the main step is Step 3 which has $\hat{i}$
iterations. According to Theorem 1 or 2, $\hat{i}$ is a $O(a)$,
i.e., $O(2^{n})$ order polynomial. Hence the network complexity of
quantum image matching is $O(2^{n})$.

Compared with the complexity of the classical image matching:
$O(2^{2n+2m})$, the quantum algorithm dropped the complexity
obviously.

\section{Discussions and conclusions}
In this paper, we try to solve the problem of measurement and give a
quantum image matching algorithm.

The contributions of this paper include:
\begin{enumerate}[1.]
\item Give a quantum image matching scheme which can find out a
small image in a big image. The entanglement property of quantum
sates make all the pixels be processed simultaneously, which
improves the scheme's effectiveness.
\item Solve the problem of measurement based on Grover's algorithm.
Our scheme can get the right answer with high probability by being
processed and measured only once, which helps to truly drop the
scheme's complexity. This is the main advantage compared with the
most existing quantum image processing algorithms.
\end{enumerate}

Future works, or something that should be discussed
include:
\begin{enumerate}[1.]
\item The number of the matching areas.

If the small image is only matched with one area in the big image
(like we discussed above), our solution is a good choice. However,
if there are no or more than one matching areas, some problems will
arise:
\begin{enumerate}[$\bullet$]
\item More than one areas.

If there are $l$ ($l>1$) matched areas, Eq. (16) will have more than
one $-\frac{1}{2}$. As a consequence, the state before measurement
will be
$$\Psi_{33}=\left(t_{\hat{i}},\cdots,t_{\hat{i}},t_{\hat{i}0},t_{\hat{i}},\cdots,t_{\hat{i}},t_{\hat{i}0},t_{\hat{i}},\cdots,t_{\hat{i}}\right)^{\text{T}},$$
Hence, one of and only one of the $l$ matched areas will be measured
randomly.

\item No area.

If there are no matched areas, in Eq. (16), all the elements of
$\Psi_{31}$ are $\frac{1}{2}$. According to Eq. (28), no matter how
many times Step 3 is iterated, the output state of Step 3 will be
$$
\Psi_{33}=\left(\frac{1}{2^{n}},\cdots,\frac{1}{2^{n}},\frac{1}{2^{n}},\frac{1}{2^{n}},\cdots,\frac{1}{2^{n}}\right)^{\text{T}}.
$$
Hence, an arbitrary pixel will be measured randomly.

\end{enumerate}

The above two points indicate that our scheme is not good at the
multi-matching-area or none-matching-area problem. This is one of
the future work.

\item Fuzzy matching.

Our scheme can only solve the problem of precisely matching. If the
images have some deformations such as shown in Fig. 4, the scheme
will fail. Hence, we plan to solve the matching problem with
deformations in the future papers.
\end{enumerate}

\begin{acknowledgements}
The authors thank Prof. Sabre Kais and Phd. Student Yudong Cao at
Purdue University for their valuable suggestions.
\end{acknowledgements}

% BibTeX users please use one of
%\bibliographystyle{spbasic}      % basic style, author-year citations
%\bibliographystyle{spmpsci}      % mathematics and physical sciences
%\bibliographystyle{spphys}       % APS-like style for physics
%\bibliography{}   % name your BibTeX data base

% Non-BibTeX users please use

\end{document}